\pdfoutput=1
\documentclass[11pt,letter,subeqn,fleqn]{article}
\usepackage[normalem]{ulem}
\usepackage{amsmath, amssymb,amsthm,cite}
\usepackage[dvips]{graphics}
\usepackage[usenames,dvipsnames]{color}
\usepackage{graphicx}
\usepackage{tabulary}
\usepackage{float}
\usepackage{gensymb}
\usepackage{authblk}
\usepackage{caption}
\usepackage{subcaption}
\usepackage{xcolor} % added by Shawn

\usepackage{mathtools}

\usepackage{epsfig}
\usepackage{latexsym}
\usepackage{amsmath}
\usepackage{amssymb}
\usepackage{mhsetup}
\usepackage{mathtools}
\usepackage{xfrac}
\numberwithin{equation}{section}
\numberwithin{table}{section}
\numberwithin{figure}{section}
\usepackage{epsfig}
\usepackage{epstopdf}
\usepackage{tabulary}
\usepackage{threeparttable}
\usepackage{array}
\usepackage{booktabs}
\usepackage{multirow}
\usepackage{pbox}
\usepackage{lipsum}
\usepackage{enumerate}
\usepackage{caption,setspace}
\usepackage[hyperfootnotes=false]{hyperref}
\hypersetup{ % citebordercolor=Violet,  filebordercolor=Red, linkbordercolor=Blue}
 colorlinks=true, %set true if you want colored links
 citecolor=ForestGreen,
 linkcolor=NavyBlue, %blue
 urlcolor=Violet}

\def\beq{\begin{equation}}
\def\eeq{\end{equation}}
\def\barr{\begin{array}{ll}}
\def\earr{\end{array}}

\def\const{\hbox{\rm const}}

{\theoremstyle{definition}

}

\newcounter{tabnum}

\title{Spatiotemporal Behaviour of SIR Models with Cross-Diffusion and Vital Dynamics}

\author[1]{Maryam Ahmadpoortorkamani\thanks{Electronic mail: qzm279@usask.ca}}
\author[2]{Alexei Cheviakov\thanks{Corresponding Author. Electronic mail: alexei.cheviakov@usask.ca}}
\affil[1,2]{Department of Mathematics and Statistics, University of Saskatchewan}
\date{}

\begin{document}

\maketitle

\begin{abstract}
Contemporary epidemiological models often involve spatial variation, providing an avenue to investigate the averaged dynamics of individual movements. In this work, we extend a recent model by Vaziry, Kolokolnikov, and Kevrekidis [Royal Society Open Science 9 (10), 2022] that included, in both infected and susceptible population dynamics equations, a cross-diffusion term with the second spatial derivative of the infected population density. Diffusion terms of this type occur, for example, in the Keller-Siegel chemotaxis model.  The presented model corresponds to local orderly commute of susceptible and infected individuals, and is shown to arise in two dimensions as a limit of a discrete process. The present contribution identifies and studies specific features of the new model's dynamics, including various types of infection waves and buffer zones protected from the infection. The model with vital dynamics additionally exhibits complex spatiotemporal behaviour that involves the generation of quasiperiodic infection waves and emergence of transient strongly heterogeneous patterns.

\end{abstract}

\section{Introduction}

In the context of epidemiological modeling, the population is commonly stratified into several compartments representing distinct states with respect to a particular infectious disease. In susceptible-infected-recovered (SIR)-type models, the compartments respectively include the susceptible (S), infected (I), and recovered (R) classes. SIR models play a fundamental role in epidemiological research by allowing to simulate the dynamics and estimate specific qualitative and quantitative characteristics of infectious diseases within a given population. The specification of interplay between compartment dynamics over time and external inputs is an essential part of modeling disease spread and understanding the impact of interventions and control measures. Common epidemiological models assume a well-mixed population, and may have additional simplifying restrictions, such as constant population size, unchanging disease transmission and removal rates, or lack of demographic dynamics (births or deaths). Kermack and McKendrick's paper that appeared in 1927 \cite{bib13} is one of the earliest and most well-known attempts to formulate an SIR model in terms of ordinary differential equations (ODE) to simulate the spread of epidemics in a population with no specific structure and a constant total number of individuals. While simple and lacking factors like age, sex, infectivity variations, or spatial considerations, this model laid a foundation for numerous subsequent epidemic models which have been expanded and adapted in various ways to better represent real-world disease dynamics.
Modern studies, including Refs.~\cite{bibintro13,bibintro14,bibintro15,bibintro16, bibintro17}, use extensively modified SIR-based approaches to align with the specific characteristics of particular diseases, including the recent COVID-19.

Factoring in spatial dependence, one gains the ability to simulate the averaged dynamics of individual movements and, consequently, to study how this mobility influences the geographic progression of the disease. Early attempts to investigate epidemic models that considered the spatial diffusion of population compartments involved versions of the Lotka-Volterra species interaction system, in which the diffusion of species was incorporated; an example is provided in by the model
\begin{equation*}
\begin{array}{l}
u_t=d_{1} u_{xx} - kuv,\\[2ex]
v_t=d_{2} v_{xx} + kvu - \lambda v
\end{array}
\end{equation*}
of Ref.~\cite{bib3}, where the compartments $u$ and $v$ depend on spatial variables, $d_1$ and $d_2$ are different diffusion coefficients,  $kuv$ denotes the infection term, and $\lambda v$ the death term. Webb \cite{bib7} proposed a similar SIR model one space dimension,
\begin{equation*}
\begin{array}{l}
S_t = D_1 S_{xx}-\beta S I, \\[2ex]
I_t = D_2 I_{xx}+\beta S I-\gamma I,\\[2ex]
R_t = D_3 R_{xx}+\gamma I
\end{array}
\end{equation*}
with $S$, $I$, and $R$ depending on $(x,t)$, $D_i$ are diffusion coefficients, $\beta$ is the rate of infection when susceptible and infected people encounter, and $\gamma$ is the removal rate for the infected population. It has been shown that for Dirichlet homogeneous boundary conditions and non-negative and continuous initial conditions, this system has a positive solution if the initial condition is positive. If diffusion coefficients are equal, the infected class dies out, and susceptible population tends to a spatially uniform steady state.

In addition to simple diffusion of individuals, spatial models can capture different contact patterns. For instance, in Ref.~\cite{bib6}, an SIR model was modified by the assumptions that the motion of susceptible individuals away from higher concentrations of infected ones took place, and that infected people moved away from overcrowded regions.

In the novel study by Vaziry, Kolokolnikov, and Kevrekidis \cite{bib2}, it was proposed that individuals are susceptible to contracting infections when they depart from their residences and temporarily move to adjacent locations. The infection process here is characterized by an absence of latency between exposure and the appearance of symptoms. The likelihood of susceptible individuals becoming infected depends on the presence of both susceptible individuals and existing infections within a specific area. It is noteworthy that individuals do not undergo random or diffusive movements upon leaving their dwellings. Rather, their mobility patterns involve a return to their original (home) location, as exemplified by routines such as shopping or work-related activities. The authors derived an SIR model within a one-dimensional spatial domain, given by
\begin{equation}\label{m7}
\begin{array}{l}
S_t =-D \beta S  I_{xx}-\beta S  I,\\[2ex]
I_t=D \beta S I_{xx}+\beta S I-\gamma I,\\[2ex]
R_t=\gamma I,
\end{array}
\end{equation}
where $D$ is the diffusion coefficient. The equations \eqref{m7} were shown to exhibit interesting behaviour including the existence of buffer zones protected from the infection. In the context of spatially homogeneous SIR models, multiple forms of $S-$, $I-$, and $R-$dependent forcing terms that can approximate various effects have been considered in the literature. In particular, vital dynamics is modeled variable or constant birth and mortality rates (e.g., Refs.~\cite{bib1, bib8, bib9, bib11, bib12} and references therein). The integration of logistic growth offers an important avenue of exploration. For example, in \cite{bib8}, an SIR model was presented where logistic growth rates were incorporated into the susceptible population, accompanied by the introduction of non-monotonic incidence and saturated treatment rates.

In this work, we begin from the derivation of a general two-dimensional PDE model
\begin{equation}\label{eq:main:model:2D:vital}
\begin{array}{l}
S_t=-D \beta S \Delta I-\beta S I  +rS-\mu S-\dfrac{r}{K} S^2,\\[2ex]
I_t=D \beta S \Delta I+\beta S I-m I,\\[2ex]
R_t=\gamma I-\mu R.
\end{array}
\end{equation}
(Section \ref{sec:2Dmodel}). Here $\Delta = \nabla^2$ is the Laplace operator, $\beta$ is the infection transition coefficient, $r$ and $K$ are the logistic growth coefficient and carrying capacity, $m$ is the total mortality rate of the infected population, and $\mu$ is the natural mortality rate. The cross-diffusion term $\sim S \Delta I$ corresponds to orderly commute of susceptible and infected individuals. In particular, the cross-diffusion terms in the PDE model \eqref{eq:main:model:2D:vital} arise exactly from a limit of the discrete process.

We note that the diffusion term in the PDE system \eqref{m7} that incorporates only the $I$-diffusion in both $S$ and $I$ evolution equations is reminiscent of that in the famous Keller-Siegel model of chemotaxis \cite{keller1971model}, given by, for example,
\begin{equation}\label{eq:KSiegel}
\begin{array}{l}
\rho_t=D_b \Delta \rho - \nabla\cdot (k \rho \nabla c) + a \rho - \beta\rho^2 ,\\[2ex]
c_t=D_c \Delta c + \alpha\rho - \gamma c.
\end{array}
\end{equation}
In \eqref{eq:KSiegel}, $\rho$ and $c$ are the bacterial density and the attractant concentration depending on $t$ and $x\in \mathbb{R}^n$, $D_b$ and $D_c$ are diffusion coefficients, $a$ is the rate of bacterial division, $\alpha$ is the rate of attractant production, $k$ denotes the chemotactic sensitivity, $\beta$ is a logistic term coefficient, and $\gamma$ is the rate of attractant decay. Indeed, in the limit of small $D_b$ and with $k\sim D_c$, the remaining diffusion terms agree with those in \eqref{eq:main:model:2D:vital}. The model \eqref{eq:KSiegel} has been extensively studied; various phenomena and solution behaviour types including, for example, shock waves, have been observed (see, e.g. \cite{wang2008shock} and references therein).

In Section \ref{sec2_1} we observe the essentially non-hyperbolic, diffusive PDE model \eqref{m7} admits solutions in the form of traveling waves of variable amplitudes, moving at different speeds, and accelerating or decelerating depending on system parameters. In Section \eqref{sec4}, Lie point symmetries of the PDE system \eqref{m7} are systematically sought. In its most general form, the PDEs admit space and time translations, and additionally, two unusual point symmetries that allow to map any solution of \eqref{m7} to a solution where the infected compartment is modified by an addition of a term proportional to an exponentially time-decaying spatially harmonic wave, of the form of Fourier separated solutions for the usual linear heat equations, but of lesser generality. This transformation is admitted despite the fact that the PDEs \eqref{m7} are neither linear nor separable. Further, for a special case of the cross-diffusion SIR model that admits additional scaling symmetries, the PDE system is reduced to a single ordinary differential equation (ODE) that yields scaling-invariant self-similar solutions. The latter correspond to infection waves propagating on a half-line when an infection source is located at the origin.

Section \ref{sec:buffer}  illustrates the existence of ``buffer zones" protected from infection by low population density in the original one-dimensional model without vital dynamics. These zones postpone infection progression, effectively acting as firebreaks.

Section \ref{sec:VDyn} studies an important property of the model \eqref{eq:main:model:2D:vital} which, in addition to cross-diffusion, involves death and logistic growth terms: the spontaneous generation of quasi-periodic infection waves that originate from a single spike of infected population in a specified location. It is shown that the initial amount of susceptible individuals controls emergence times of the infection waves. Buffer zones also arise in the model with vital dynamics; beyond such zones, infection waves propagate after a significant delay. In case of homogeneous initial and homogeneous Neumann boundary conditions, the model reduces to time-dependent ODEs that support spiral sink-type  equilibria dependent parameters of the logistic term. For large carrying capacity, the dynamics remains close to the isoline of the approximately  conserved Hamiltonian-type integral for prolonged times.

When initialized by a non-homogeneous initial condition, the cross-diffusion model with vital dynamics can exhibit striking behaviour, where the dynamics is initially unstable, leading to the appearance of ``dark spikes" (self-produced transient buffer zones) in the susceptible compartment.
It is shown that the emergence of such spikes can be attributed to linear instability of high wave number perturbations of the nontrivial steady state.

In Section \ref{sec:VDyn} we also numerically study solutions of an ODE describing time-independent states of the PDE model with cross-diffusion and vital dynamics. It is shown that there exist spatially periodic, non-harmonic positive definite equilibrium solutions.

In Section \ref{sec:2D:study} it is demonstrated that buffer zones and quasi-periodic, essentially two-dimensional infection wave generation are also features of the full PDE model \eqref{eq:main:model:2D:vital} in two spatial dimensions.

The paper is concluded with a discussion in Section \ref{sec:disc}.

%================================================================================

\section{Derivation} \label{sec:2Dmodel}

\begin{figure}[H]
\begin{center}

\includegraphics[height=6cm]{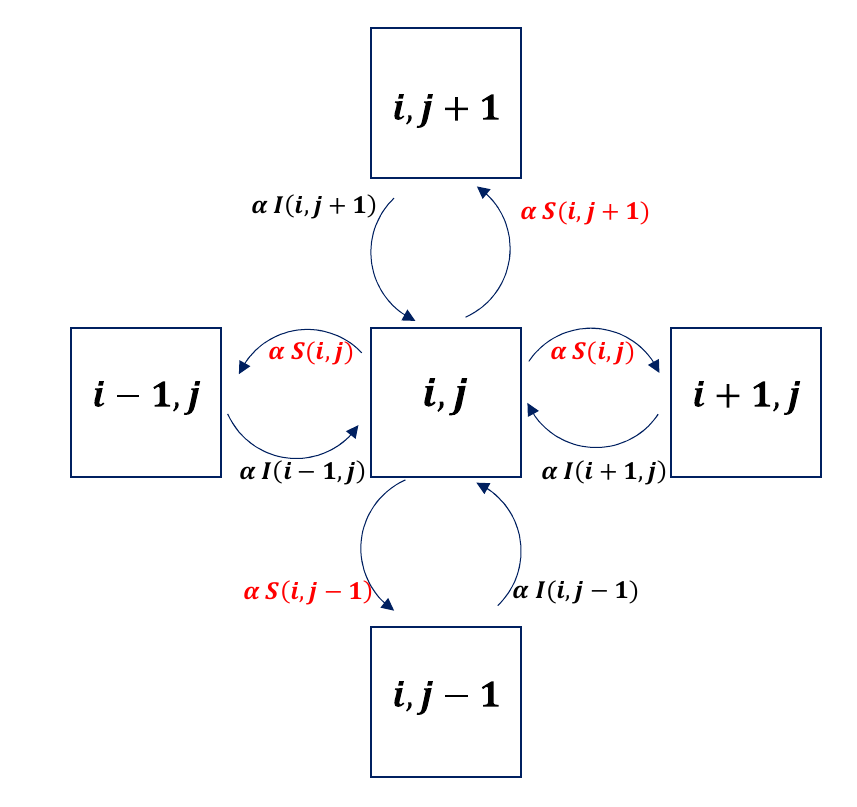}

\end{center}
\caption{\footnotesize{Mobility patterns in a lattice-based SIR model. The chart visualizes the mobility dynamics within a lattice-based SIR model, showing the flow of individuals between neighboring compartments.}}
\label{pic4}
\end{figure}

In this section, we derive the two-dimensional PDE model \eqref{eq:main:model:2D:vital} featuring the cross-diffusion terms. To characterize movements of individuals in two dimensions, consider a discrete spatial domain with bins indexed by $i$ and $j$ ranging from 1 to $N$ (Figure \ref{pic4}). These bins are utilized to represent the populations of susceptible, infected, and recovered individuals, denoted as $S(t,i,j)$, $I(y,i,j)$, and $R(t,i,j)$, respectively. Consistently with the conventional SIR model, it is assumed that infection can occur with a probability $\beta$ when a susceptible individual encounters an infected individual. Furthermore, $\alpha$ is employed to denote the travel rate, with the assumption of identical travel rates for both vertical and horizontal motions to simplify the model. Lastly, $\bigtriangleup I(t,i,j)$ denotes the new infections within bin $(i,j)$ at $[t,t+1]$.

According to Figure \ref{pic4}, the change $\delta I$ at the cell with indices $(i,j)$ is given by
\[
\barr
\delta I(t, i, j) =   \beta[ \underbrace{S(t, i, j) - 4\alpha S(t, i, j)}_{\text{available } S \text{ at }(i,j)}]\times\\[2ex]
\hspace{1cm} [\underbrace{ I(t, i, j) - 4\alpha I(i, j) + \alpha(I(t, i + 1, j) + I(t, i - 1, j) + I(t, i, j - 1) + I(t, i, j + 1)) }_{\text{available } I \text{ at }(i,j)}]\\[2ex]
+\beta(\alpha  \underbrace{S(t, i, j)}_{\text{commuter } \text{ to }(i+1,j)})\times\\  \hspace{1cm}[\underbrace{ I(t, i + 1, j) - 4\alpha I(t, i + 1, j) + \alpha(I(t, i + 2, j) + I(t, i, j) + I(t, i + 1, j - 1) + I(t, i + 1, j + 1)) }_{\text{available } I \text{ at }(i+1,j)}]\\[2ex]
+\beta(\alpha  \underbrace{S(t, i, j)}_{\text{commuter } \text{ to }(i-1,j)})\times\\\hspace{1cm}[\underbrace{I(t, i - 1, j) - 4\alpha I(i - 1, j) + \alpha(I(t, i - 2, j) + I(t, i, j) + I(t, i - 1, j - 1) + I(t, i - 1, j + 1)) }_{\text{available } I \text{ at }(i-1,j)}]\\[2ex]
+ \beta(\alpha\underbrace{S(t, i, j)}_{\text{commuter } \text{ to }(i,j+1)})\times\\[2ex]
\hspace{1cm}[\underbrace {I(t, i, j + 1) - 4\alpha I(i, j + 1) + \alpha(I(t, i - 1, j + 1) + I(t, i + 1, j + 1) + I(t, i, j + 2) + I(t, i, j))}_{\text{available } I \text{ at }(i,j+1)}] \\[2ex]
+ \beta(\alpha \underbrace{S(t, i, j)}_{\text{ commuter}  \text{ to }(i,j-1)})\times\\[2ex]
\hspace{1cm}[\underbrace {I(t, i, j - 1) - 4\alpha I(i, j - 1) + \alpha(I(t, i - 1, j - 1) + I(t, i + 1, j - 1) + I(t, i, j) + I(t, i, j - 2))}_{\text{available } I \text{ at }(i,j-1)}].
\earr
\]
The corresponding discrete-time SIR model on a lattice is expressed as follows:
\[
\barr
S(t+1,i,j)=S(t,i, j)-\delta I(t, i, j),\\[2ex]
I(t+1,i,j)=I(t,i, j)+\delta I(t, i, j)-\gamma I(t,i, j), \\[2ex]
R(t+1,i,j)=R(t,i,j)+\gamma I(t,i, j),
\earr
\]
where $\gamma$ denotes the recovery rate. Considering the grid spacing in both the $x$ and $y$ directions as $dx$ and $dy$, and defining $I(t,i, j)$ as $I(t,x,y)$ with $x=i dx$ and $y=j dy$, we can employ a Taylor expansion to obtain the following result:
\[
\barr
\delta I(t, x, y) =\beta S(t,x, y) \left(2 \alpha dx^2  I_{xx}(t,x, y) + 2 \alpha dy^2 I_{yy}(t,x, y) + I(t,x,y)\right)\\[2ex]
\qquad \qquad +O(dx^3)+O(dy^3).
\earr
\]
Under the assumption of $dx=dy$, this simplifies to
\[
\barr
 \delta I(t, x, y) =\beta D S(t,x, y) \left( I_{xx}(t,x, y) + I_{yy}(t,x, y)\right)+\beta S(t,x, y)I(t,x,y)\\[2ex]
 \qquad\qquad +O(dx^3),
\earr
\]
where $D=2 \alpha dx^2$.

We now apply the two-point forward-difference formula to $S$, leading to $S(t+1,x,y)-S(t,x,y)=S_t(t,x,y)$ (similarly for $I$ and $R$). In the limit of large $\alpha$ and small $dx$ such that $D=2 \alpha dx^2=\const$, the resulting equations become
\begin{equation}\label{m8}
\begin{array}{l}
S_t=-D \beta S( I_{xx}+I_{yy})-\beta S I,\\[2ex]
I_t=D \beta S (I_{xx}+I_{yy})+\beta S I-\gamma I,\\[2ex]
R_t=\gamma I,
\end{array}
\end{equation}
where  $t \geq 0$, $(x,y) \in \mathcal{A}=[x_a,x_b] \times [y_a,y_b] \subseteq \mathbb{R}^{2}$.

The resulting model \eqref{m8} expands upon the model \eqref{m7} to a 2D spatial domain, incorporating the movement of infected individuals in the second dimension ($y$) and their interaction with $S$. This addition to the model enables the observation of its influence on new infection dynamics.
We note that the system \eqref{m8} has a conserved quantity
\[
S(x,y,t) + I(x,y,t) + R(x,y,t) = N(x,y),
\]
where the total population, denoted by $N(x,y)$, remains constant over time. The first equation in the system enforces $S(x,y,t) \geq 0$. This condition is met since $S_0 \geq 0$, and from the first equation in \eqref{m8}, the logarithmic rate of change
$
{dS(x,y,t)}/{S(x,y,t)}= \left(-  D \beta \Delta I(x,y,t)-\beta I(x,y,t)\right) dt<0,
$
which yields, for a known solution $I=I(x,y,t)$,
\[
S(x,y,t)=S_{0}\, \exp\left(\int_{0}^{t} -\big( D \beta  \Delta I(x,y,\tau)+ \beta I(x,y, \tau)\big)\, d\tau\right) > 0.
\]
In a similar manner, the second and the third PDEs in \eqref{m8} ensure the positivity of $I$ and $R$.

We now incorporate vital dynamics in the form of logistic growth for the susceptible population, $S_t\sim rS\left(1-{S}/{K}\right)$, where $r$ denotes the intrinsic growth rate, and $K$ represents the carrying capacity arising from internal factors such as competition among susceptible individuals. We introduce a constant natural mortality rate $\mu$ for all model compartments. Additionally, we consider $\gamma$ to be the recovery rate, and $\omega$ the extra death rate for infected individuals, signifying the fatality of the disease. As a whole,
\beq\label{eq:vital:m}
m=\mu+\gamma+\omega
\eeq
corresponds to the removal rate for the infected population. In two spatial dimensions, this extension yields the PDE system \eqref{eq:main:model:2D:vital}, which completes the derivation.

\section{Waves in the SIR model with cross-diffusion} \label{sec2_1}

The spatiotemporal PDE model \eqref{m7} supports various kinds of waves that behave in a stable manner. In the current section we discuss two instances of wave behaviour, the first involving the degeneration of an initial infection spike, and the second describing a self-similar exact solution corresponding to an incoming infection wave.

\subsection{Accelerating and decelerating waves}\label{eq:sec:accel:decel}

\begin{figure}[H]
\begin{center}

\includegraphics[width=0.5\textwidth]{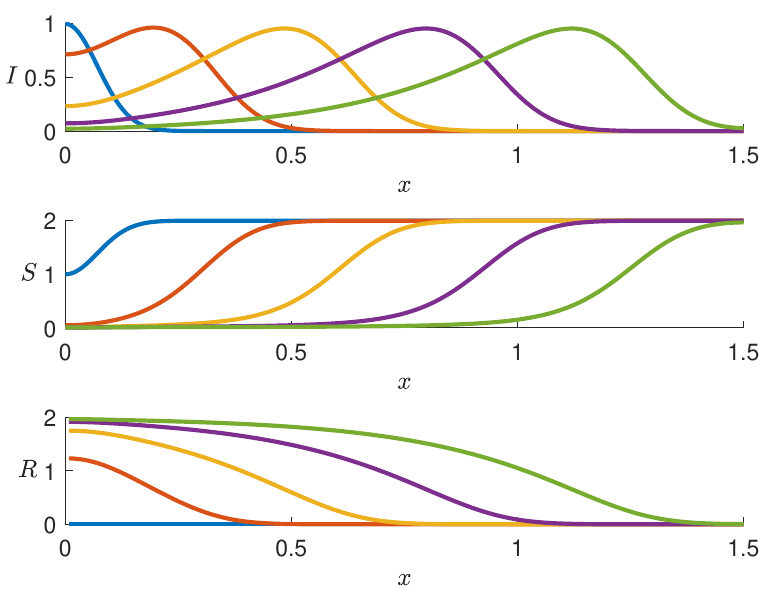}

\end{center}
\caption{\footnotesize{ Dynamics of the SIR Model \eqref{m7}: wave shapes of $I$, $S$, and $R$ at $t=0, 3, 6, 9, 12$.}}
\label{fig:2:1}
\end{figure}

Consider an initial value problem for the nonstandard diffusion PDE system \eqref{m7} involving a one-dimensional spatiotemporal domain $x\in [-L, L] \subseteq \mathbb{R}$, $t\geq 0$. Suppose that initially, $I$ is given by a peak located at the origin, with initial amplitude $A$ and characteristic wavelength $\lambda$, $R=0$, the total population $N$, and $S=N-I$. The corresponding wave degeneration leads to a traveling wave whose speed of propagation depends on $A$, $\lambda$, and the diffusion coefficient $D$, and the parameters $\beta$ and $\gamma$ of the model. From the dimensional considerations, noting the physical dimensions $[D]=\text{m}^2$, $[\beta] = [\gamma]=\text{s}^{-1}$, $[A] = [\lambda]=\text{m}$, the wave speed $[v]=m\text{s}^{-1}$, and the time $[t]=s$,  one can construct a system of invariants and use the Buckingham $\pi$-theorem (see, e.g., Ref.~\cite{bibintro30}) to express the wave speed as a function of the other characteristics,
\beq\label{eq:v:of:rest}
v = D^{1/2} \beta \,\mathcal{F} (\beta\gamma^{-1}, A\lambda^{-1}, D\lambda^{-2}, \beta t),
\eeq
where $\mathcal{F}$ is a certain unknown function. It can be shown numerically that depending on the parameters, the PDE system \eqref{m7} supports both accelerating and decelerating waves. As an example, we perform a dimensionless computation fixing $\beta=1$, $\gamma=0.4$, $A=1$, $\lambda\sim 1$ by choosing the initial condition $I(x,0) = \exp(-100\, x ^2)$,  $S(x,0) = 2 - I(x,0)$, $R_0 = 0$, and varying $D$. For $D=10^{-3}$, the propagation of infection and recovery waves is illustrated in Fig. \ref{fig:2:1}.

\begin{figure}[H]
\begin{center}
\includegraphics[width=0.4\textwidth]{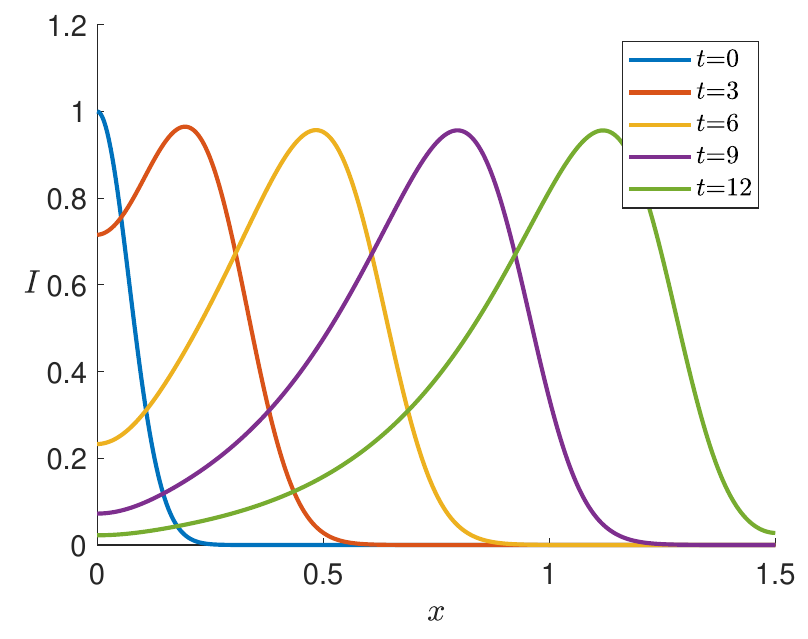} \quad \includegraphics[width=0.4\textwidth]{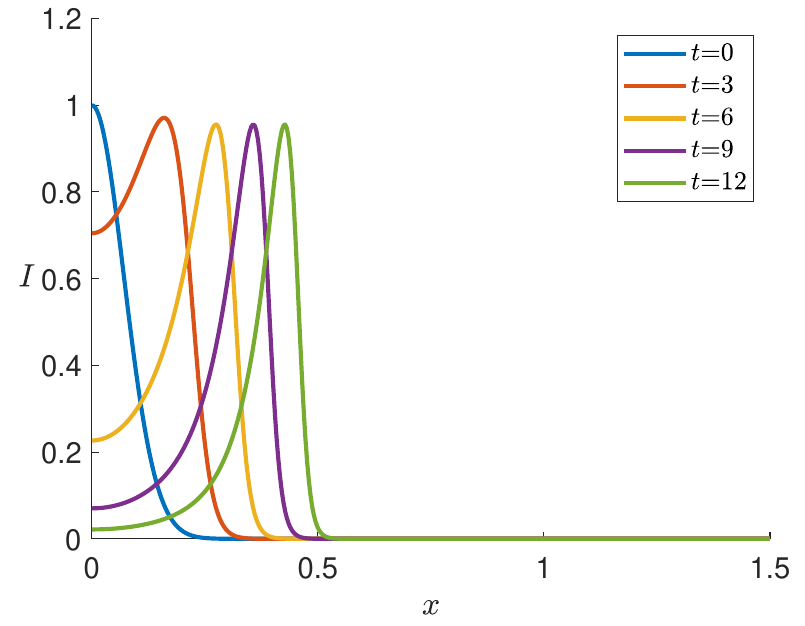}
\end{center}
\caption{\footnotesize{Influence of the diffusion coefficient on the infection wave dynamics: $D=10^{-3}$ (left), $D=10^{-5}$ (right).}}
\label{fig:2:2}
\end{figure}

\begin{figure}[H]
\begin{center}
\includegraphics[width=0.5\textwidth]{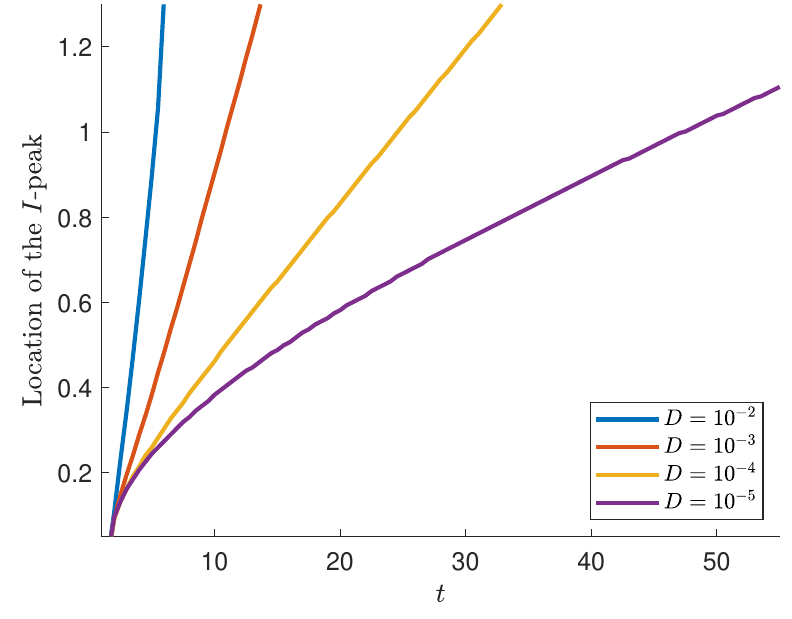} \hspace{1cm}
\end{center}
\caption{\footnotesize{The $I-$peak dynamics in the SIR model \eqref{m7} and its variation with $D$.}}
\label{fig:2:3}
\end{figure}

It can be shown that the speed of the propagation of the infection wave in this setup significantly depends on the value of $D$. Figure \ref{fig:2:2} shows that, as it is intuitively expected, higher diffusion coefficients correspond to higher wave speeds and higher wavelengths. A comparison of $I-$wave peak dynamics in Figure \ref{fig:2:3} shows increasing and decreasing peak velocity values for a range of diffusion coefficients. To quantify this behaviour, the peak trajectories obtained in Figure \ref{fig:2:3} can be fit into the power law $X(t;D)=A(D) t^{P(D)}$ using least squares; the resulting values for $A(D)$ and $P(D)$ are given in Table \ref{Tab1}. The corresponding peak velocity estimates $V(t;D)=A(D)P(D) t^{P(D)-1}$ follow therefrom.

\begin{table}[h]
\centering
\caption{\footnotesize{ Estimates $A(D)$ and $P(D)$ for the infection peak trajectories fit into $X(t;D)=A(D) t^{P(D)}$ for different $D$}}
\begin{tabular}{ccc}
\toprule
Diffusion coefficient $D$ &  $P(D)$ &  $A(D)$ \\
\midrule
$0.01$ & $1.9330$ & $0.04060$\\
$10^{-3}$ & $1.2531 $&$0.0501$ \\
$10^{-4}$ & $0.8624$& $ 0.0635$ \\
$10^{-5}$ & $0.6437$&$0.0843$ \\
\bottomrule
\end{tabular}
\label{Tab1}
\end{table}

\subsection{Symmetries and self-similar infection waves}\label{sec4}

We now calculate point symmetries of the PDE system \eqref{m7} in order to seek physically meaningful symmetry-invariant solutions.\footnote{See, e.g., Ref.~\cite{bibintro29} and references therein for details.}

In the general case, the PDE system \eqref{m7} admits the following point symmetry generators:
\beq\label{eq:main:symm}
X_1 = \dfrac{\partial}{\partial x}, \quad X_2 = \dfrac{\partial}{\partial t},\quad
X_3 = \cos \left(\dfrac{\sqrt{\beta}x}{\sqrt{a}}  \right)e^{-\gamma t}\dfrac{\partial}{\partial I},\quad X_4 = \sin \left(\dfrac{\sqrt{\beta}x}{\sqrt{a}}  \right)e^{-\gamma t}\dfrac{\partial}{\partial I}.
\eeq
Here $X_1$ and $X_2$ correspond to space and time translations, and $X_3$ and $X_4$ are interesting symmetries that add a time-decaying, spatially oscillatory part to any solution of \eqref{m7} (here $a=D\beta$). For example,
the global transformation corresponding to $X_3$ is given by
\begin{equation}\label{eq:symm:transfX3}
x^*=x,\quad t^*=t,\quad S^*(x^*,t^*)=S(x,t), \quad I^*(x^*,t^*)=I(x,t)+C \left ( \cos \left(\dfrac{\sqrt{\beta}x}{\sqrt{a}}  \right)e^{-\gamma t} \right),
\end{equation}
where $C$ is an arbitrary constant. In particular, if $S(x,t)$ and $I(x,t)$ are parts of any solution of the system of PDEs \eqref{m7}, then  $S^*(x,t)$ and $I^*(x,t)$ represent the corresponding components of a new solution.

The limited set of symmetries full PDE system \eqref{m7} does not include scalings. However, it is easy to see that a scaling-invariant version of \eqref{m7} can be obtained in a certain limit. Indeed, consider a special class of models where $\gamma, \beta \ll 1$ while $D \beta = a \sim 1$; it is given by a coupled PDE system\footnote{A similar kind of re-scaling was used in Ref.~\cite{wang2008shock} where shock waves for the chemotaxis model were obtained.}
\begin{equation}\label{m17}
\begin{array}{l}
\dfrac{dS}{dt}=-a  S(x,t) I_{xx}(x,t),\\[2ex]
\dfrac{dI}{dt}=a  S(x,t) I_{xx}(x,t)\\
\end{array}
\end{equation}
The dynamics of $R$ still follows the third equation of \eqref{m7} but is decoupled from the two PDEs \eqref{m17} and may be omitted. The system \eqref{m17} may be naturally considered in the domain $x,t>0$, with certain initial and boundary conditions
\beq\label{eq:ICs:symm}
I(x,t=0) = I_0(x),\quad  S(x,t=0) = N - I_0(x),\quad  I(x=0,t) = A, \quad  I(x \rightarrow \infty,t) = B,
\eeq
where the total population $N$ assumed constant for the moment. The PDE system \eqref{m17} admits six point symmetries with generators
\beq\label{eq:SSxx:symms}
\barr
X_1 = \dfrac{\partial}{\partial I}, \qquad X_2 = \dfrac{\partial}{\partial x}, \qquad X_3 = \dfrac{\partial}{\partial t},\\[2ex]
X_4 = x \dfrac{\partial}{\partial I}, \qquad X_5 = 2t \dfrac{\partial}{\partial t} + x \dfrac{\partial}{\partial x}, \qquad X_6 = I \dfrac{\partial}{\partial I} + S \dfrac{\partial}{\partial S} - t \dfrac{\partial}{\partial t}.
\earr
\eeq
One can obtain exact self-similar solutions upon reduction of the PDE system \eqref{m17} with respect to the scaling symmetry $X_5$. Constructing and solving the characteristic equation, we get
\begin{equation}\label{m19}
I(x,t)=f(z), \qquad S(x,t)=g(z),
\end{equation}
where $f$ and $g$ are so far unknown functions of the invariant $z = {x}/{\sqrt{t}}$. It is easy to see that the initial and boundary conditions \eqref{eq:ICs:symm} can be accommodated by the invariant ansatz \eqref{m19}:
\begin{equation}\label{m18}
f(0)=I(x=0,t)=A, \qquad f(z\to \infty)=I(x,t=0)=I(x\to \infty ,t)=B,
\end{equation}
where $A$ and $B$ are arbitrary non-negative constants. The substitution of \eqref{m19} into the PDEs \eqref{m17} naturally yields $g(z) = \const -f(z) = N - f(z)$ and a single second-order ODE for $f(z)$ given by
\begin{equation}\label{m21}
2a (N-f)f''+z f'=0,
\end{equation}
with the boundary conditions $f(0)=A$, $f(\infty)=B$, or alternatively, initial conditions $f(0)=A$, $f'(0)=h(A,B)$, where $h$ is chosen so that $f$ satisfies $f(\infty)=B$.

\begin{figure}[H]
\begin{center}
\includegraphics[height=5cm]{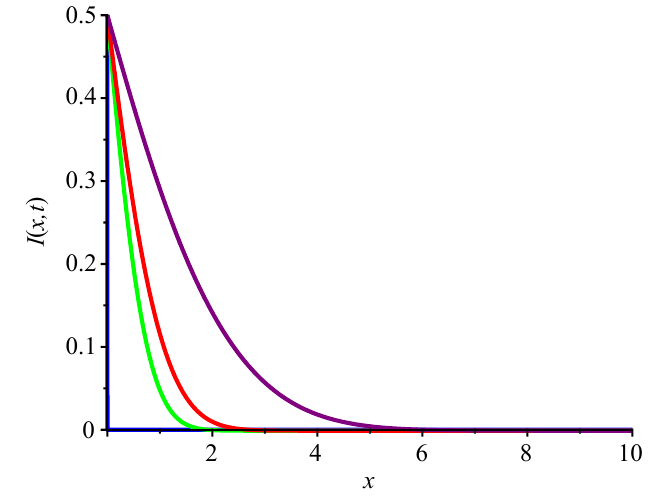} \qquad \includegraphics[height=5cm]{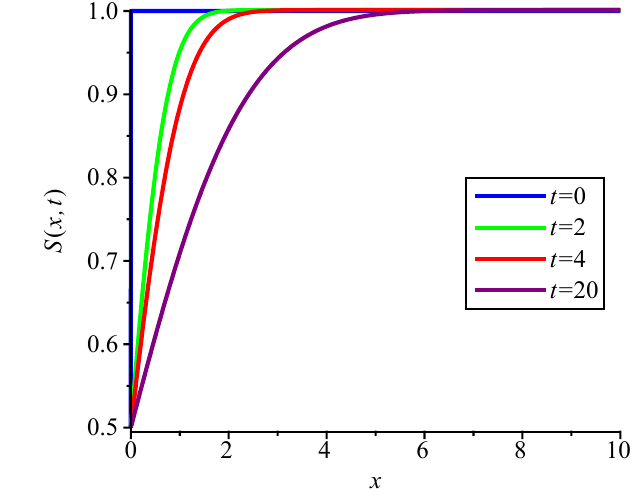}
%\hspace{1cm}
\end{center}
\caption{A self-similar solution of \eqref{m17} modeling an expanding infection wave in half-space.}
\label{fig:2:4}
\end{figure}

A sample numerical solution of the ODE \eqref{m21} corresponding to $N=1$, $A=0.5$ and $B=0$ has $f'(0)=-1$. The dynamics of the infected and susceptible populations $I(x,t)=f({x}/{\sqrt{t}})$, $S(x,t)=g({x}/{\sqrt{t}})$ is shown in Figure \ref{fig:2:4}, where the infected population expands across the spatial domain, monotonously increasing at every spatial location. The boundary condition corresponds to the infection source at the origin.

\subsection{Buffer zones}\label{sec:buffer}

An important feature of the SIR PDE model with nonstandard diffusion \eqref{m7} is the existence of situations that prevent the transmission of infections within a spatial \emph{buffer zone}, a region that consistently remains uninfected and exhibits a natural resistance to the disease.

For small values of the diffusion coefficient, the susceptible population follows $S_t \sim -\beta S  I$ and therefore is a monotone decreasing function: for an initial condition $S(x,0)=S_0(x)$, one has $S(x,t)\leq S_0(x)$. The dynamics of the infected population is approximately governed by the ODE
\[
\dfrac{I_{t}}{I}  \simeq \beta S-\gamma \lesssim \beta S_{0}-\gamma.
\]
The function $I(x,t)\sim I_0(x) \exp((\beta S_{0}-\gamma)t)$ therefore decreases approximately exponentially when $\beta S_{0}-\gamma<0$. Note that the parameter $R_0(x) = {\beta}S_0/{\gamma}$ is the basic dimensionless reproduction number. It is defined as the expected number of secondary cases produced by a single (typical) infection in a completely susceptible population at any $x$ in spatial domain. The condition $R_0(x)<1$ thus corresponds to local exponential decay of $I$.

\begin{figure}[H]
\begin{center}
\includegraphics[width=0.6\textwidth]{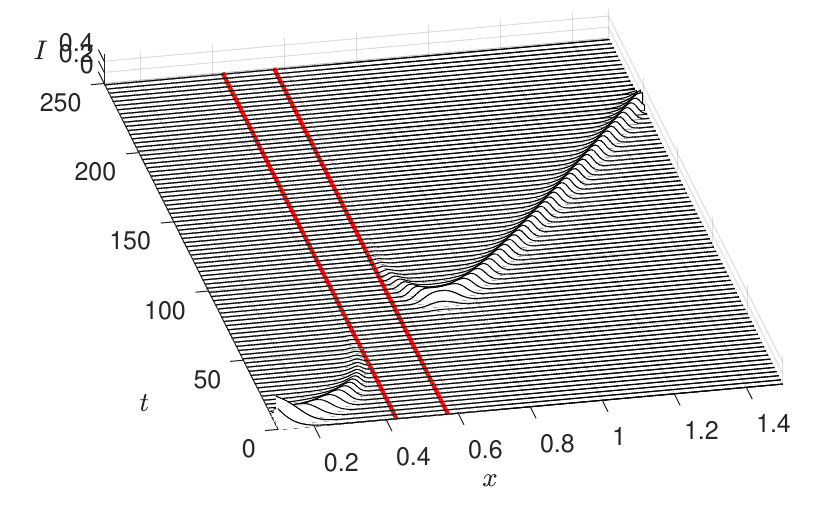}
\end{center}
\caption{$I(x,t)$ in a buffer zone (between the red lines).}
\label{fig:buffer:1D}
\end{figure}

\begin{figure}[H]
\begin{center}
\includegraphics[width=0.7\textwidth]{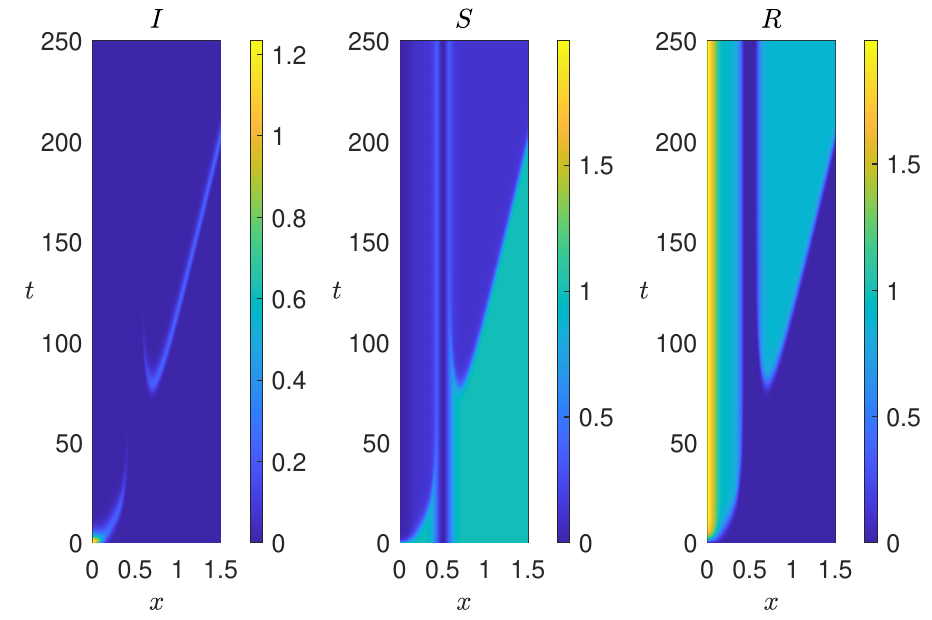}
\end{center}
\caption{$I(x,t)$, $S(x,t)$, and $R(x,t)$ in a buffer zone.}
\label{fig:buffer:1D2}
\end{figure}

An example of a buffer zone is given in Figures \ref{fig:buffer:1D}, \ref{fig:buffer:1D2} where $x\in [0, 1.5]$, $D=10^{-5}$, the initial infected population is concentrated around the origin, $I(x,0)=\exp(-100x^2)$, the susceptible population is small around $x=0.5$, $S(x,0)=1-\exp(-100(x-0.5)^2)$,
and homogeneous Neumann boundary conditions are imposed. with $\beta=1$ and $\gamma=0.4$ the initial reproduction number is negative in the neighbourhood of $x=0.5$. It is observed that the infection wave re-emerges on the other side of the buffer zone, starting to grow substantially from the spatial points around $x\sim 0.7$ where the susceptible population is sufficient.

%=========================================================================
\section{Spatio-temporal SIR model with vital dynamics} \label{sec:VDyn}

In the case of vital dynamics being present, in the form of logistic growth for the susceptible population, constant natural mortality rate $\mu$ for all compartments, a nonzero recovery rate, and an extra death rate for infected individuals, the corresponding model is given by \eqref{eq:main:model:2D:vital}. In one spatial dimension, the PDE system is given by
\begin{equation}\label{eq:1D:with:vital}
\begin{array}{l}
S_t=-D \beta S I_{xx}-\beta S I  +rS-\mu S-\dfrac{r}{K} S^2,\\[2ex]
I_t=D \beta S I_{xx}+\beta S I-m I,\\[2ex]
R_t=\gamma I-\mu R.
\end{array}
\end{equation}
This model has several important characteristic features related specifically to both the nonstandard diffusion operator structure and the presence of vital dynamics. Similar effects in two spatial dimensions are discussed in Section \ref{sec:2D:study} below.

\subsection{Quasi-periodic infection waves, infection delays, and buffer zones}\label{sec3_1}

\begin{figure}[H]
  \centering
  \includegraphics[width=0.7\textwidth]{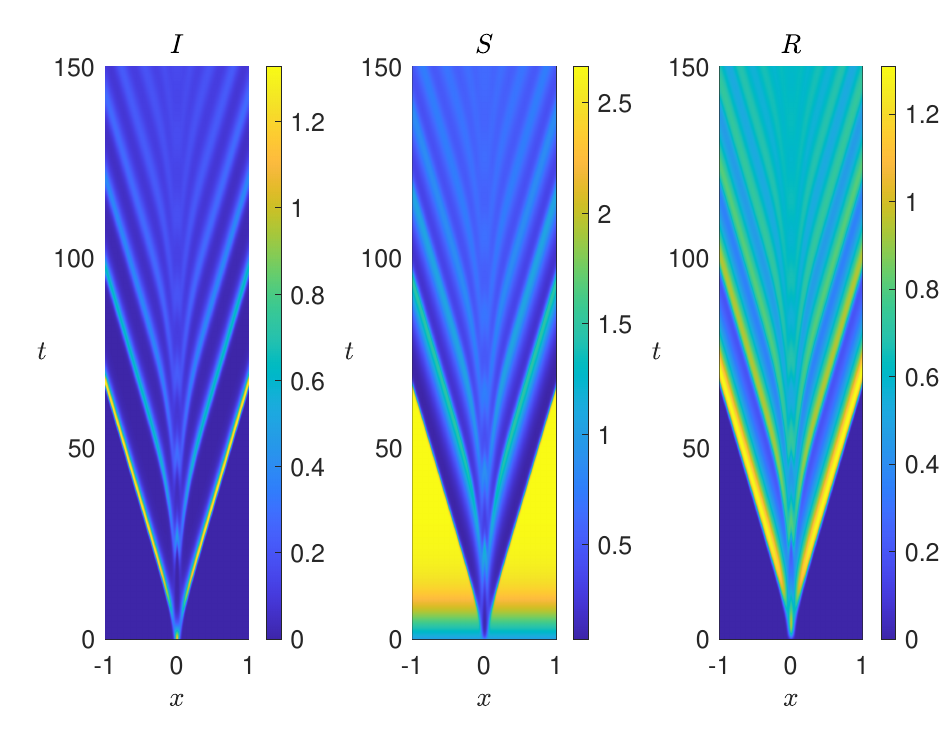}
\caption{Dynamics of the three compartments in model \eqref{eq:1D:with:vital} with $D = 10^{-5}$, $\beta = 1$, $\gamma = 0.4$, $\mu = 0.1$,  $m = 0.9$, $K = 4$, $r = 0.3$, $I_0 = \exp(-1000 \cdot x^2)$, $S_0 = 1$, $R_0 = 0$.}
\label{fig:3:1}
\end{figure}

\begin{figure}[H]
\centering
\includegraphics[width=0.5\textwidth]{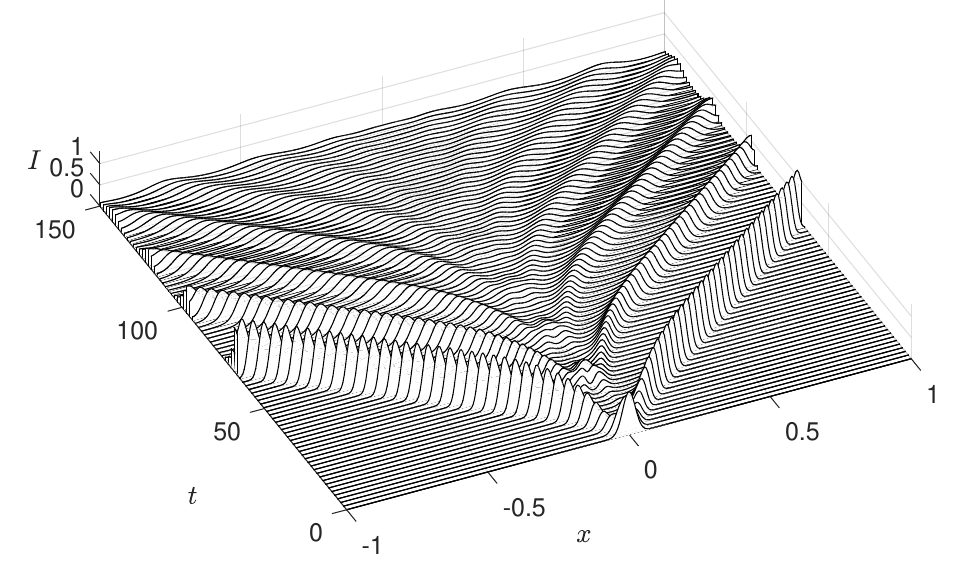} \includegraphics[width=0.46\textwidth]{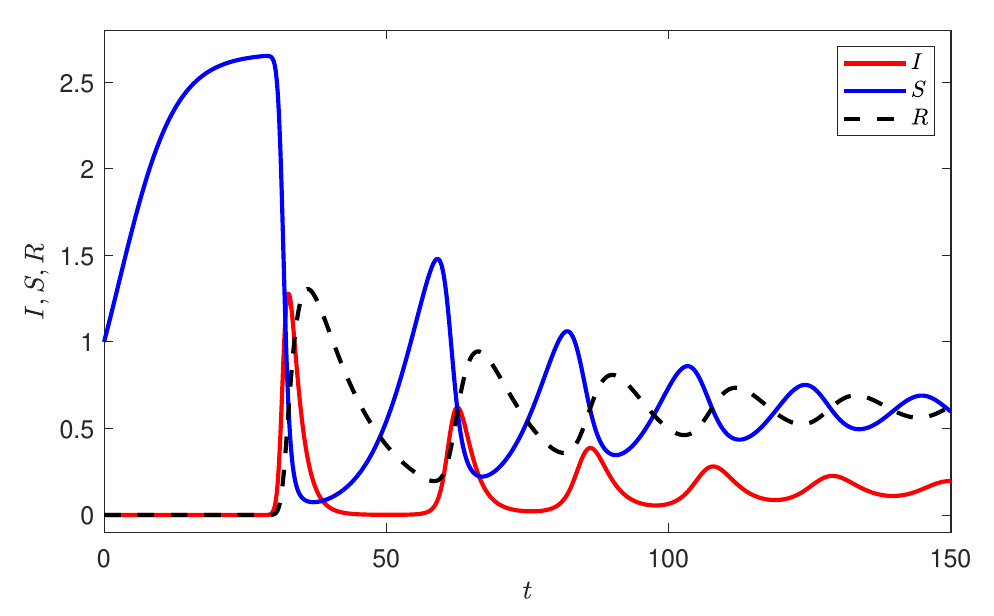}
\caption{Left: the dynamics of the infected population waves. Right: $I$, $S$, and $R$ compartments at $x=1/9$.}
\label{fig:3:2}
\end{figure}

For an interval $x\in [-L,L] \subseteq \mathbb{R}$, consider an initial population of infected individuals localized about $x=0$. As before, the system following \eqref{eq:1D:with:vital} demonstrates disintegration of the peak in the form of waves traveling to the left and to the right (Section \ref{eq:sec:accel:decel}), the key difference, however, is that for $r>\mu$, the $S-$group is not aged or exposed to a high-risk environment, the population of susceptible individuals experiences initially exponential growth. The susceptible population can re-grow in the neighborhood of $x=0$ and cause a subsequent infection wave. A numerical solution for a sample parameter set is shown in Figures \ref{fig:3:1} and \ref{fig:3:2}.
Figure \ref{fig:3:1} depicts the dynamics of \eqref{eq:1D:with:vital} within a 1-D spatial domain over an extended time period. In this representation, infection originates from $x=0$ and gradually propagates both to the left and right sides over time. As infection spreads, susceptible individuals are converted into infected ones. For instance, at $x=0.5$, where the number of infected individuals was nearly zero at $t=0$, the susceptible population increases as long as infection has not reached that location. However, once the infection reaches that point, the density of susceptible people decreases. This process repeats over time as the population of susceptible individuals oscillates. Figure \ref{fig:3:2} provides a visual representation of the spread of infected individuals across the interval $-1 < x < 1$ starting from $x=0$. This graph illustrates the recurring cycle of infection spreading outwards from its origin.

%\begin{equation}\label{m14}
%\begin{split}
% D &= 10^{-5}, \quad \gamma = 0.4, \quad \beta = 1, \quad \mu = 0.1, \quad m = 0.9, \quad K = 4, \quad r = 0.3, \\[1ex]
%I_0 &= \exp(-1000 \cdot (x-1)^2), \quad S_0 = 1, \quad R_0 = 0,
%\end{split}
%\end{equation}
%where $x\in [0,2]$ is the spatial domain.

\begin{figure}[H]
\begin{center}
\includegraphics[width=0.9\textwidth]{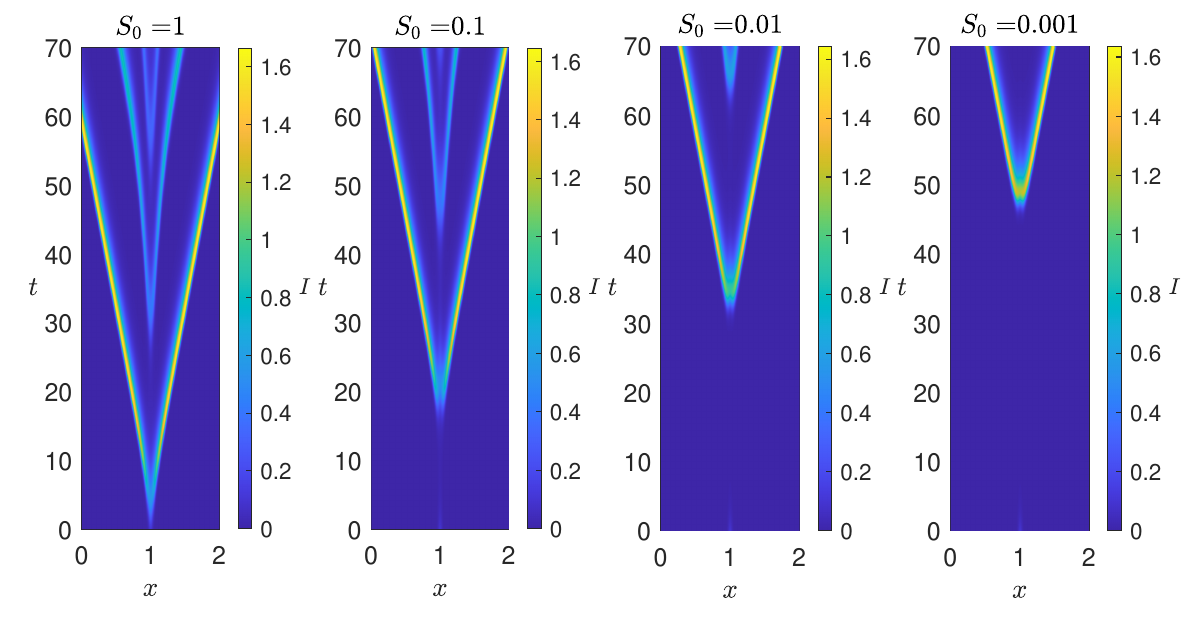}
\end{center}
\caption{{Waiting period for infection propagation: the effect of the initial susceptible population size $S_0$. }}
\label{fig:3:3}
\end{figure}

It is easy to observe that when the initial susceptible population is not large enough to sustain the infection growth, there may be a waiting period for the infection to start propagating, and the duration of the delay is dependent on the initial value of $S_0$. An illustration is provided in Figure \ref{fig:3:3} where all parameters are the same as in Figure \ref{fig:3:1} except for different $S_0$ values of $1$, $0.1$, $0.01$, and $0.001$.

\begin{figure}[H]
\begin{center}
\includegraphics[width=0.7\textwidth]{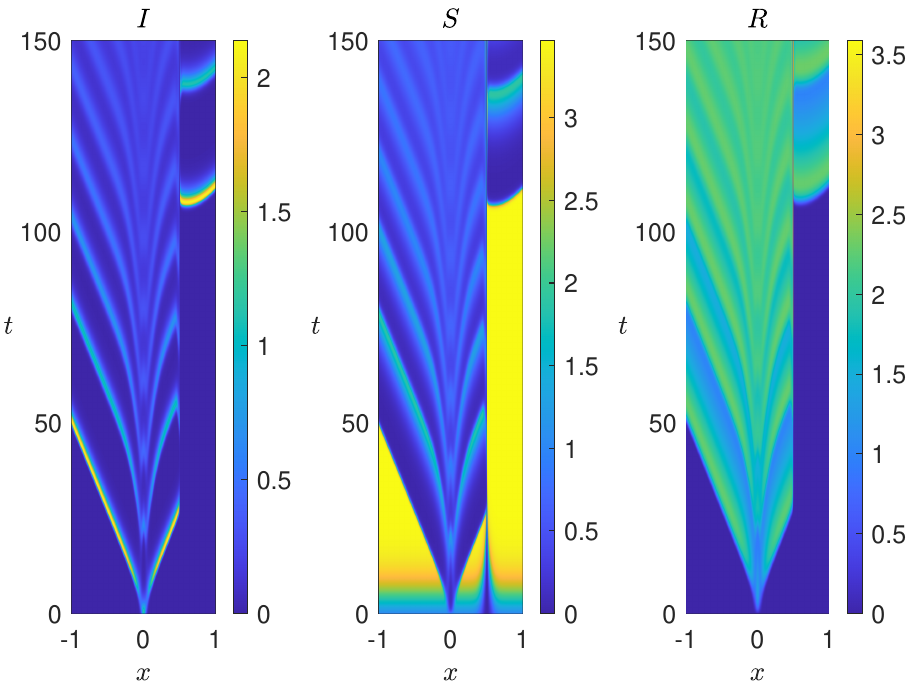}
\end{center}
\caption{{Dynamics of the three compartments in model \eqref{eq:1D:with:vital} with vital dynamics: the case of a buffer zone. }}
\label{fig:3:4}
\end{figure}

The PDE model with vital dynamics \eqref{eq:1D:with:vital} can also exhibit the buffer zone behaviour similar to the original model \eqref{m7}. Buffer zones occur in domains where, assuming small diffusion effects, the local reproduction number $R= {\beta} S/m$ satisfies $R<1$ causing the decay of the infected population. A sample situation of that kind is illustrated in Figure \ref{fig:3:4}. The infection waves originating from $x=0$ are initially blocked by the buffer zone located at $x=0.5$, but subsequently re-emerge beyond it. In this computation, $0\leq x \leq 1.5$, $D = 10^{-5}$, $\beta = 1$, $\gamma = 0.4$, $\mu = 0.04$,  $\omega=0.1$, $K = 4$, $r = 0.3$, $I_0 = \exp(-1000 \cdot x^2)$, $S_0 = 1-\exp(-100 \cdot (x-0.5)^2)$, $R_0 = 0$.

\subsection{Quasiperiodicity in the SIR model with vital dynamics}\label{sec3_2}

We are now interested in investigating close-to-equilibrium states within the framework of the model \eqref{eq:1D:with:vital}. We consider the first two equations of the system, since the class $R$ is decoupled and does not influence the dynamics of the epidemic. In the limit $D\to 0$, the model \eqref{eq:1D:with:vital} with $m=\mu+\gamma+\omega$ \eqref{eq:vital:m} becomes an ODE system
\begin{equation}\label{m1:ode}
\begin{array}{l}
S_t=-\beta S I  +rS-\mu S-\dfrac{r}{K} S^2,\\[2ex]
I_t=\beta S I-m I,
\end{array}
\end{equation}
and the pairs $(I_{\rm triv}, S_{\rm triv})=(0,0)$ and
\beq\label{eq:ss:star}
I_e=\dfrac{1}{\beta}\left(r-\mu-\dfrac{rm}{K \beta}\right), \qquad S_e=\dfrac{m}{\beta}
\eeq
represent equilibrium states of the system. The Jacobian matrix of the nontrivial equilibrium is given by
\[
J(I=I_e, S=S_e)={\left[{\begin{array}{cc}
 -\dfrac{rm}{K \beta} & -m\\[2ex]
r-\mu-\dfrac{rm}{\beta K} & 0
\end{array}}\right]}.
\]
The trace of $J$ is negative and the determinant is positive, which indicates the linear stability of the equilibrium point $(I_e, S_e)$. Moreover, when $rm /(K \beta) = 0$, it corresponds to a center-type steady state. Considering a scenario where $K\gg 1$, and $rm / (K \beta)\ll 1$ becomes negligible and examining the eigenvalues of the matrix
\[
J\left(I_e=\dfrac{1}{\beta}(r-\mu)\geq0, S_e=\dfrac{m}{\beta}\right)={\left[{\begin{array}{cc}
0 & -m\\
r-\mu & 0 \\
\end{array}}\right]},
\]
one has $\lambda_{1,2} = \pm i \sqrt{ m(r-\mu)} $. These imaginary eigenvalues do not represent hyperbolic equilibrium points since their real parts are zero, and hence the Hartman-Grobman theorem does not apply to this nontrivial steady state.

We now numerically investigate the behavior of the nonlinear PDE system \eqref{eq:1D:with:vital} in a one-dimensional spatial domain, focusing on the conditions of large $K$, near the nontrivial steady state. For the initial condition, it is imperative that the system is initialized in close proximity to its spatially constant nontrivial steady state $S=S_e$, $I=I_e$ \eqref{eq:ss:star}; it follows that for homogeneous Neumann boundary conditions, the dynamics is determined by the ODE \eqref{m1:ode}. As a particular example, consider the model \eqref{eq:1D:with:vital} in a 1D spatial domain $[0, 1.5]$ with the parameter values
\beq\label{m13}
\gamma=0.4,\quad \beta=1,\quad \mu=\omega=0, \quad r = 0.4,
\eeq
and the initial conditions
\beq\label{m13:ICs}
I_0 = I_e + 0.3,\quad S_0 = S_e +0.3, \quad R_0 = 0.
\eeq

\begin{figure}[H]
\begin{center}
\includegraphics[width=0.53\textwidth]{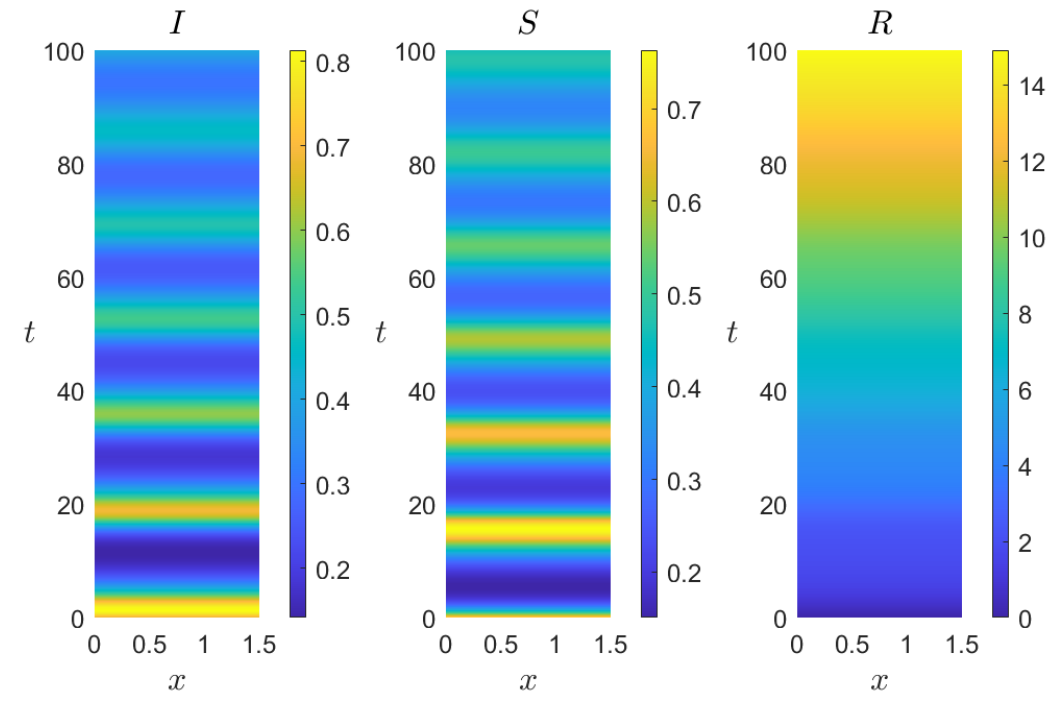}~\includegraphics[width=0.45\textwidth]{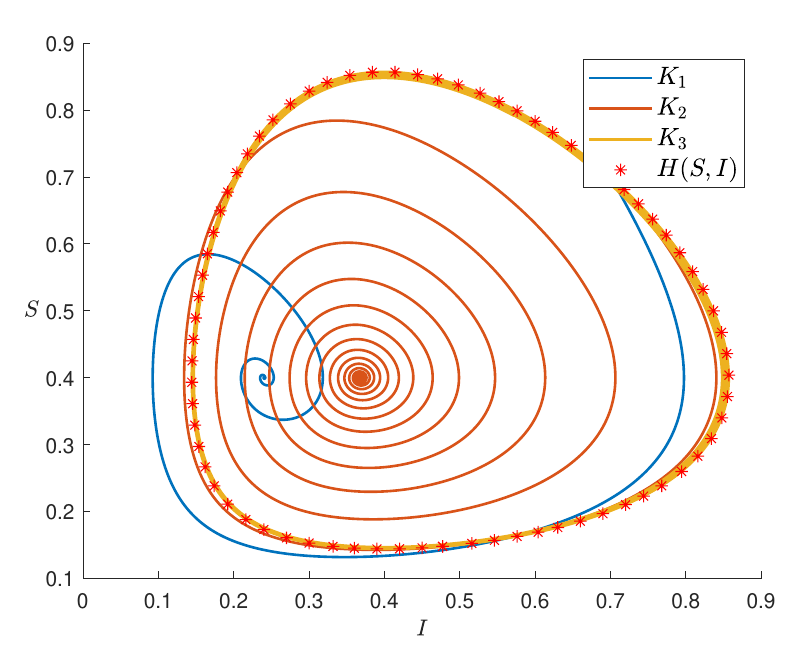}
\end{center}
\caption{Left: solution of the PDE  \eqref{eq:1D:with:vital} with \eqref{m13}, \eqref{m13:ICs} and $K=5$. Right: the phase plane trajectory of the corresponding system \eqref{m1:ode} for $K=(1, 5, 10^4)$, parameters \eqref{m12}, and initial conditions \eqref{m13:ICs} corresponding to $K=10^4$. Asterisks denote the level curve of the approximately conserved integral \eqref{m12}.}
\label{fig:spiral}
\end{figure}

Figure \ref{fig:spiral} (left) corresponds to $K=5$ and illustrates spatially homogeneous population distribution decaying in time due to the logistic term. Figure \ref{fig:spiral} (right) shows the $(I,S)$ phase plane of the system for the three values of the logistic coefficient $K=(1, 5, 10^4)$. For large $K$, the system dynamics approaches the Hamiltonian ODE given by \eqref{m1:ode} with $K\to\infty$ that has closed phase trajectory
that conserves the Hamiltonian
\begin{equation}\label{m12}
H(S,I)=\beta I-(r-\mu)\ln I  + \beta S-m \ln S  =\const
\end{equation}
(a well-known exact integral for the Lotka-Volterra-type systems). Indeed, for $\overline{S}=\ln S$ and $\overline{I}=\ln I$, the corresponding ODE system
\begin{equation}\label{m4}
\begin{array}{l}
{\overline{S_t}=-\beta e^{\overline{I}}+r-\mu},\\[2ex]
{\overline{I_t}=\beta e^{\overline{S}}-m}
\end{array}
\end{equation}
satisfies the Hamiltonian equations
\[
\frac{d\overline{S}}{dt}=-H_{\overline{I}}, \qquad \frac{d\overline{I}}{dt}=H_{\overline{S}},
\]
for
\[
H(\overline{S},\overline{I})=\beta e^{\overline{I}}-(r-\mu)\overline{I} + \beta e^{\overline{S}}-m\overline{S}=\const,
\]
the latter being equivalent to \eqref{m12}.

As illustrated in Figure \ref{fig:spiral} (right), for $K\sim 1$, the logistic term leads to the system with a spiral sink at the equilibrium $I_e(K)$, $S_e(K)$ \eqref{eq:ss:star}, whereas for $K\gg1$, the trajectory approaches the curve \eqref{m12}.

\subsection{Dark spike formation and quasi-equilibria}\label{sec:equil:tr}

An interesting feature of quasi-equilibria that emerge for smaller values of the carrying capacity $K$ and larger diffusion coefficients $D$ is the formation of ``dark spikes" that effectively correspond to buffer zones, or low-value ``valleys" in the susceptible population. As a specific example, let
\begin{subequations}\label{dark_spike_init_data}
\beq\label{dark_spike_consts}
D =0.01,\quad \beta=1,\quad \gamma=0.4,\quad \mu=0.1,\quad K=4, \quad r = 0.4,\quad \omega=0.2,% \quad m=0.7,
\eeq
with periodic initial conditions related to equilibrium values \eqref{eq:ss:star} as follows:
\beq
I_0 = I_e, \quad S_0 = S_e ( 1+ A\cos(\pi n x/L)), \quad R_0 = 0,
\eeq
\end{subequations}
where $x\in [0,L]$, $L=1.5$, $A=0.5$ is the harmonic perturbation amplitude, and $n=6$ is the mode number.

\begin{figure}[H]
\begin{center}
\includegraphics[width=0.8\textwidth]{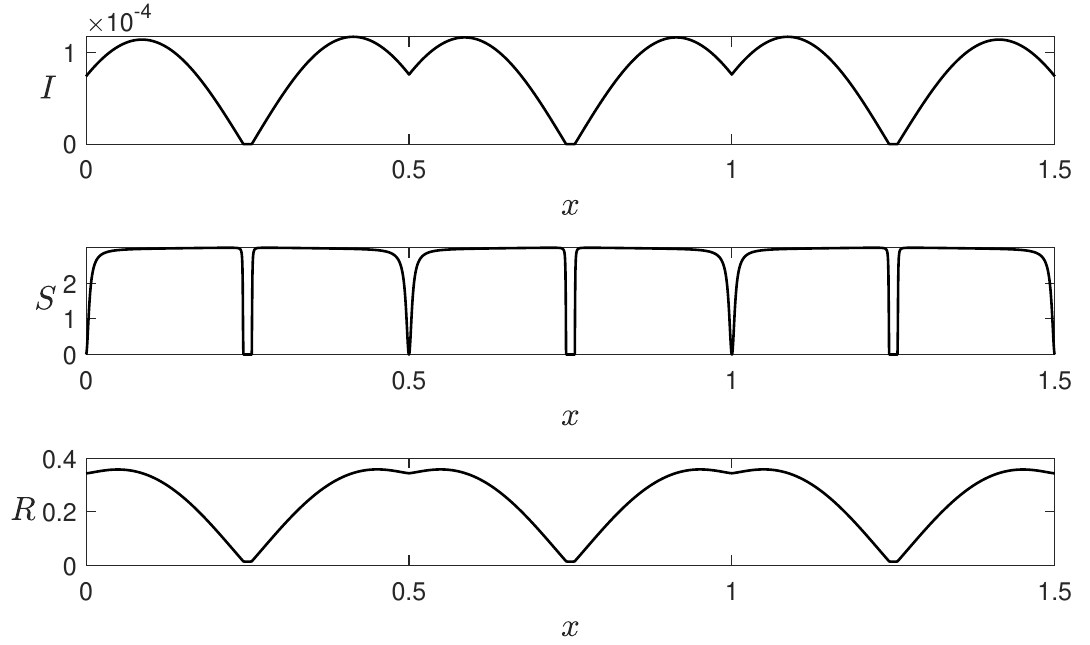} \hspace{1cm}
\end{center}
\caption{Cross-section of Figure \ref{pic15} at $t=55$.}
\label{pic15b}
\end{figure}

Figure \ref{pic15} illustrates the behavior of the full PDE system \eqref{eq:1D:with:vital} for all points $x$ in the spatial domain. In particular, the formation of buffer zones in the form of ```dark spikes" is observed in $S$. It has been shown that this effect is not related to numerical error as the same qualitative and quantitative behaviour is observed on different spatial and temporal meshes. A spatial cross-section at $t=55$ is shown in Figure \ref{pic15b}. Interestingly, lower values of recovery rate $\gamma$ cause further local instability. For example, choosing $\gamma=0.1$ and other parameters as in \eqref{dark_spike_consts}, one gets a set of forming, and then eventually vanishing, double-peak dark spikes (Figure

\begin{figure}[H]
\begin{center}
\includegraphics[width=0.8\textwidth]{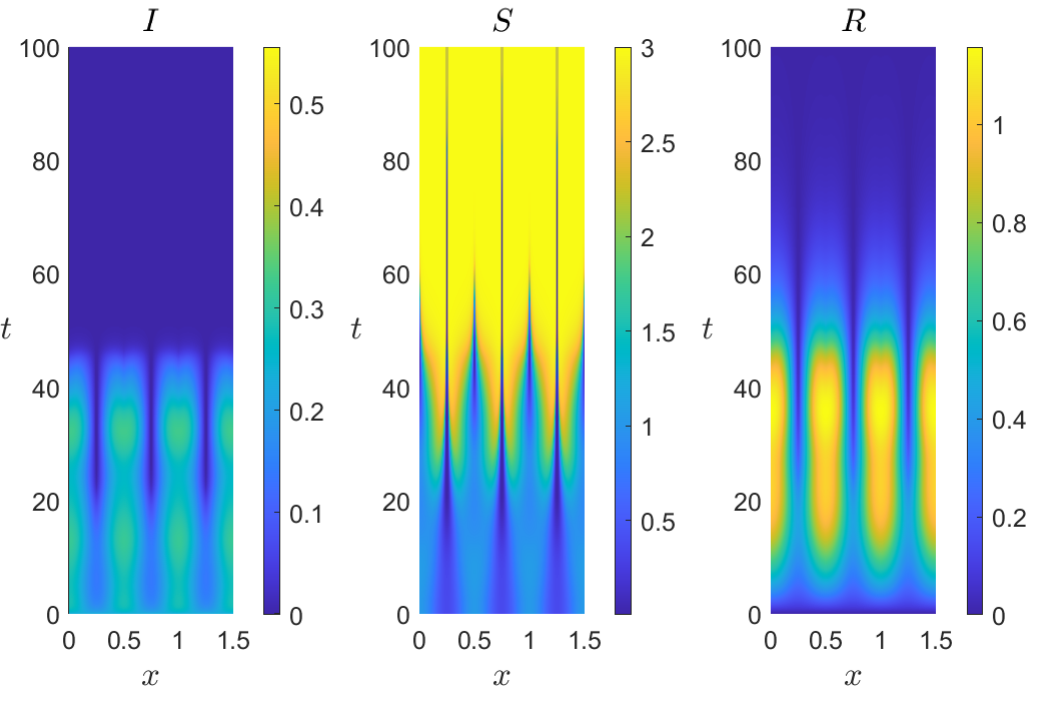} \hspace{1cm}
\end{center}
\caption{{``Dark spike" formation in the PDE system \eqref{eq:1D:with:vital} with \eqref{dark_spike_init_data}.}}
\label{pic15}
\end{figure}
%%%%%%%%%%%%%%

%%%%%%%%%%%%%%

\begin{figure}[H]
\begin{center}
\includegraphics[width=0.8\textwidth]{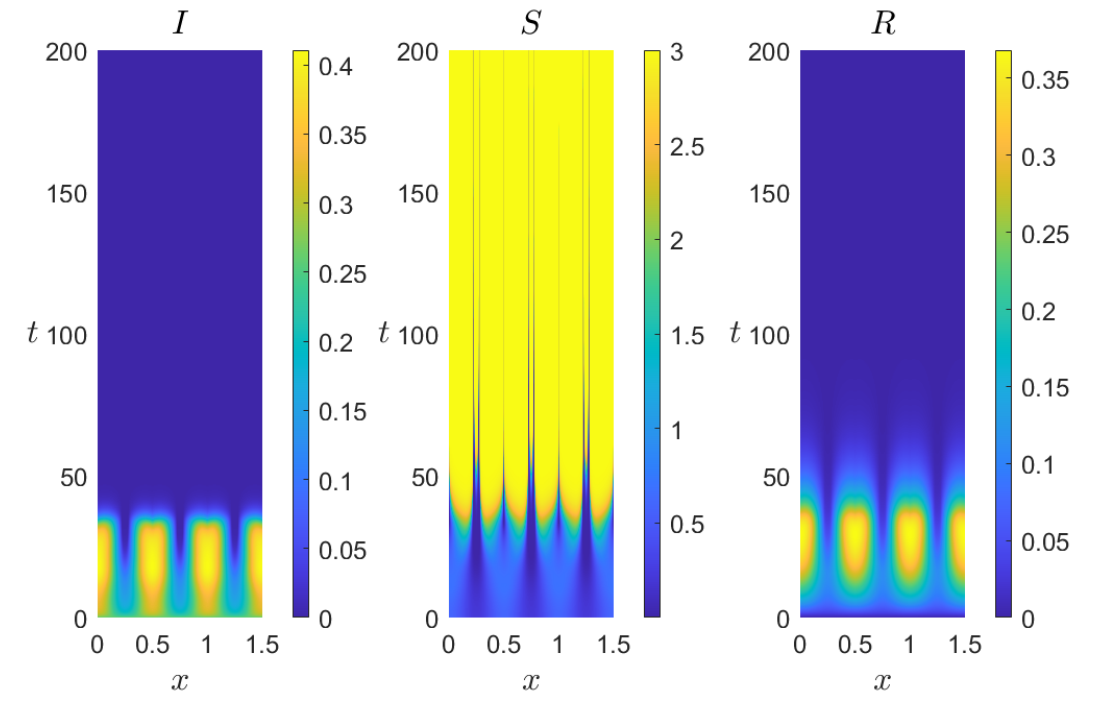} \hspace{1cm}
\end{center}
\caption{{Double ``dark spikes" in case of a lower recovery rate $\gamma=0.1$.}}
\label{pic15}
\end{figure}
%%%%%%%%%%%%%%

\begin{figure}[H]
\begin{center}

\includegraphics[width=0.46\textwidth]{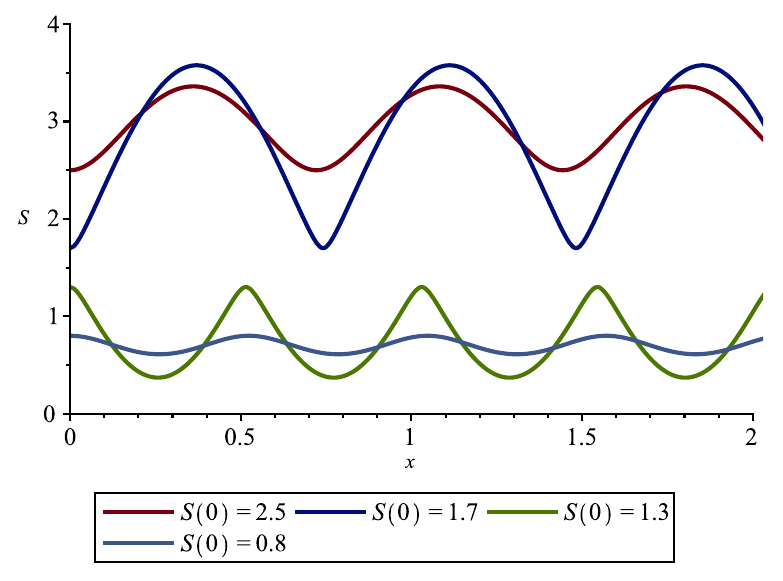} ~~~~~
\includegraphics[width=0.46\textwidth]{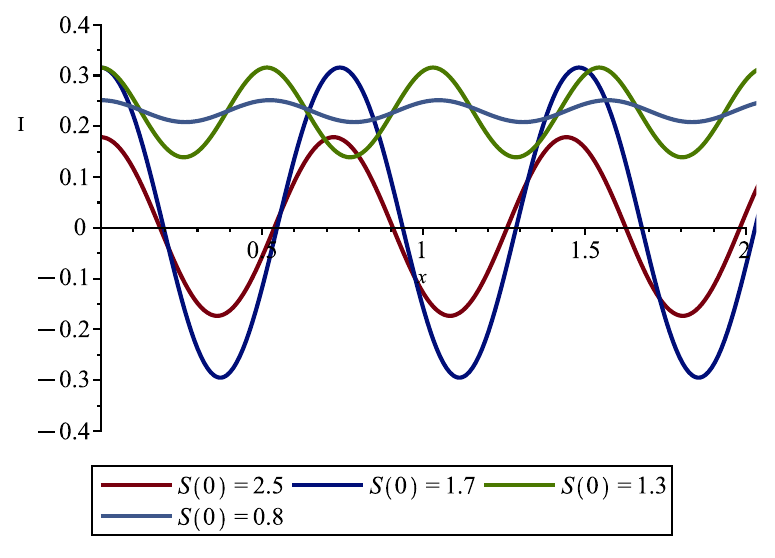}
\end{center}
\caption{Physical (nonnegative $\tilde{I}$, $\tilde{S}$) and non-physical (negative $\tilde{I}$) solutions of the time-independent ODEs \eqref{eq:1D:with:vital:ODEs}}. Left: curves of $\tilde{S}$ solving \eqref{eq:vit:ODE:S} for different initial conditions; right: the corresponding curves of $\tilde{I}(x)$
\label{fig:ODE:solsS}
\end{figure}

The formation of quasi-equilibrium states, in particular, dark spike formation, is related to solution forms of the time-independent version of the system \eqref{eq:1D:with:vital} given by
\begin{equation}\label{eq:1D:with:vital:ODEs}
\begin{array}{l}
D \beta \tilde{S} \tilde{I}''= - \beta \tilde{S} \tilde{I}  +(r-\mu) \tilde{S}-\dfrac{r}{K} \tilde{S}^2 ,\\[1ex]
- D \beta \tilde{S} \tilde{I}'' = \beta \tilde{S} \tilde{I}-m \tilde{I},\\[1ex]
\gamma \tilde{I}=\mu \tilde{R}.
\end{array}
\end{equation}
where the $x$-dependent equilibrium states are denoted by $\tilde{S}=\tilde{S}(x)$, $\tilde{I}=\tilde{I}(x)$, and $\tilde{I}=\tilde{I}(x)$, and primes denote $x$-derivatives.
Adding the first two equations, one has $m \tilde{I} = (r-\mu) S-({r}/{K}) S^2$, and thus the equilibrium states $I(x)$ and $R(x)$ are expressed through $S(x)$:
\beq\label{eqil:x:I:R}
\tilde{I} = \dfrac{1}{m}\left((r-\mu) \tilde{S}-\dfrac{r}{K} \tilde{S}^2\right),\qquad \tilde{R} = \dfrac{\gamma}{\mu}\tilde{I},
\eeq
As a result, $\tilde{S}$ satisfies the ODE obtained by substituting \eqref{eqil:x:I:R} into the first equation of \eqref{eq:1D:with:vital:ODEs}:
\beq\label{eq:vit:ODE:S}
D {\beta}  (K(\mu - r) + 2r\tilde{S}) \tilde{S}'' +  2D\beta  r(\tilde{S}')^2  + \beta r \tilde{S}^2 + ({\beta} K (\mu - r) - m r )\tilde{S} - K m (\mu - r)=0.
\eeq
Solutions $\tilde{S}(x)$ of the differential equation \eqref{eq:vit:ODE:S} are in general not expressed in terms of elementary functions. It can be shown numerically that the ODE \eqref{eq:vit:ODE:S} admits non-harmonic periodic solutions that can satisfy homogeneous Neumann or periodic boundary conditions (Figure \ref{fig:ODE:solsS}). Some of such solutions yield positive values of $I(x)$, while others may yield negative values.

The emergence of ``dark spikes" is related to the linear instability of the nontrivial steady state \eqref{eq:ss:star} of the PDE system  \eqref{eq:1D:with:vital}. Indeed, consider its linear harmonic perturbation
\beq\label{eq:vit:equil:harmpert}
I = I_e + \epsilon I_1 e^{i(kx-\omega t)},\qquad S = S_e + \epsilon S_1 e^{i(kx-\omega t)},
\eeq
$\epsilon \ll 1$. The substitution of \eqref{eq:vit:equil:harmpert} into \eqref{eq:1D:with:vital} and retention of $O(\epsilon)$ terms only yields the amplitude relationship
\[
I_1 = -\left(\dfrac{\mu - r}{Dmk^2} + \dfrac{r}{K\beta D k^2}\right) S_1
\]
and the dispersion relation
\beq \label{eq:disprel}
\omega = -\dfrac{i}{K\beta D k^2} Q, \qquad Q= (K\beta (\mu-r) + 2 m r ) D k^2  - (K\beta (mu-r) + m r).
\eeq
In particular, $\omega$ is imaginary for all $k$, which yields decay of the corresponding perturbation mode \eqref{eq:vit:equil:harmpert} when $Q>0$, and exponential growth when $Q<0$. For example, when the general mortality exceeds the recovery rate, $\mu>r$,  the two cases correspond to $k^2 > k_0^2$ and $k^2 < k_0^2$, respectively, with
\[
k_0^2 = \dfrac{ K\beta (\mu-r) + m r}{K\beta (\mu-r) + 2 m r}\,.
\]
Consequently, modes of larger wavelengths $\lambda>\lambda_0 = 2 \pi/k_0$ are unstable. By contrast, in the example \eqref{dark_spike_consts} where dark spikes arise, one has $\mu<r$, and the quantity $Q$ in \eqref{eq:disprel} has the form $Q=-0.0064 k^2 + 0.92$, which is negative, and corresponding modes are unstable, for $k^2>143.75$. The formation of dark spikes can thus be attributed to an attempted linear blowup of small-wavelength modes, compensated by nonlinear effects.

%exponential growth or exponential decay of the harmonic perturbations \eqref{eq:vit:equil:harmpert} based on the sign on the bracket in the right hand side of \eqref{eq:disprel}.

%%%%%%%%%%%%%%

%=========================================================================
\section{The SIR model with cross-diffusion in two dimensions} \label{sec:2D:study}

\begin{figure}[H]
\begin{center}
\includegraphics[width=0.9\textwidth]{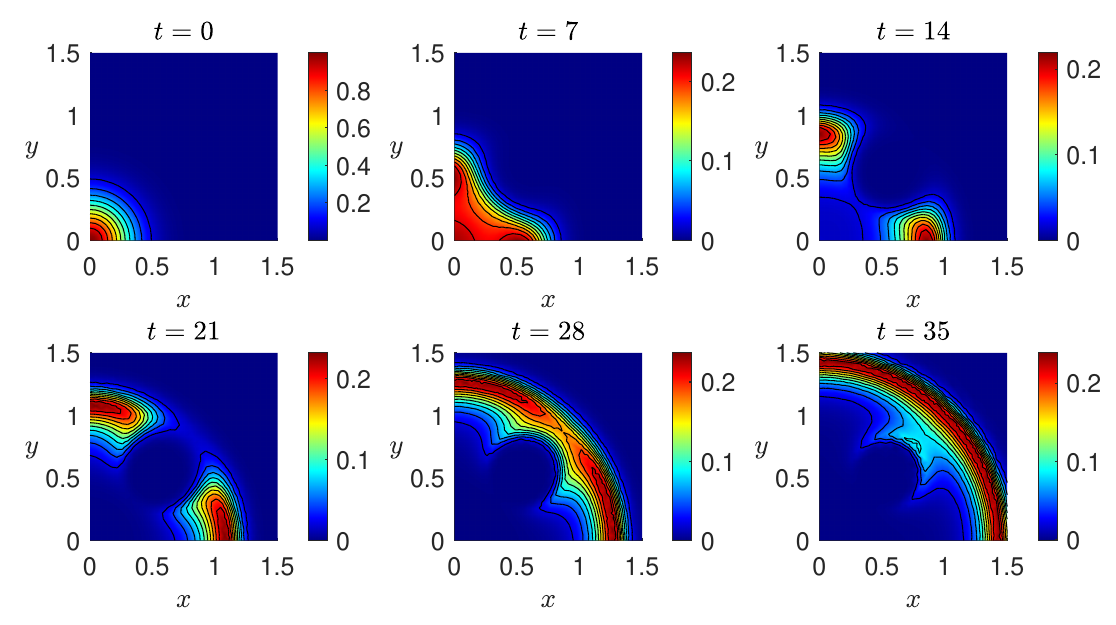}
\end{center}
\caption{ A buffer zone for the two-dimensional PDE model \eqref{m8} with cross-diffusion and no vital dynamics.}
\label{fig:2d:buffer}
\end{figure}

The two-dimensional PDE model with cross-diffusion \eqref{eq:main:model:2D:vital} is capable of producing types of behaviour similar to those for the one-dimensional model. We first illustrate the existence of buffer zones. As an example, consider a spatial domain $[0, 1.5]\times [0, 1.5]$, and the PDE model \eqref{m8} with Neumann boundary conditions and the following parameter values and initial conditions:
\begin{equation}\label{m10}
\begin{split}
&D=10^{-5}, \quad \gamma=0.4, \quad \beta=1,\\[1ex]
&I(x, y, 0) = \exp\left(-10\left(x^2 + y^2\right)\right),\\[1ex]
&S(x, y, 0) = 1 - \exp\left(-10\left((x-0.5)^2 + (y-0.5)^2\right)\right),\\[1ex]
&R(x, y, 0) = 0,
\end{split}
\end{equation}
such that the low susceptible density region is centered at $M=(0.5,0.5)$. In this case, a circular region around $M$ is avoided by the infection for all times (Figure \ref{fig:2d:buffer}).

\begin{figure}[H]
\begin{center}

\includegraphics[width=0.9\textwidth]{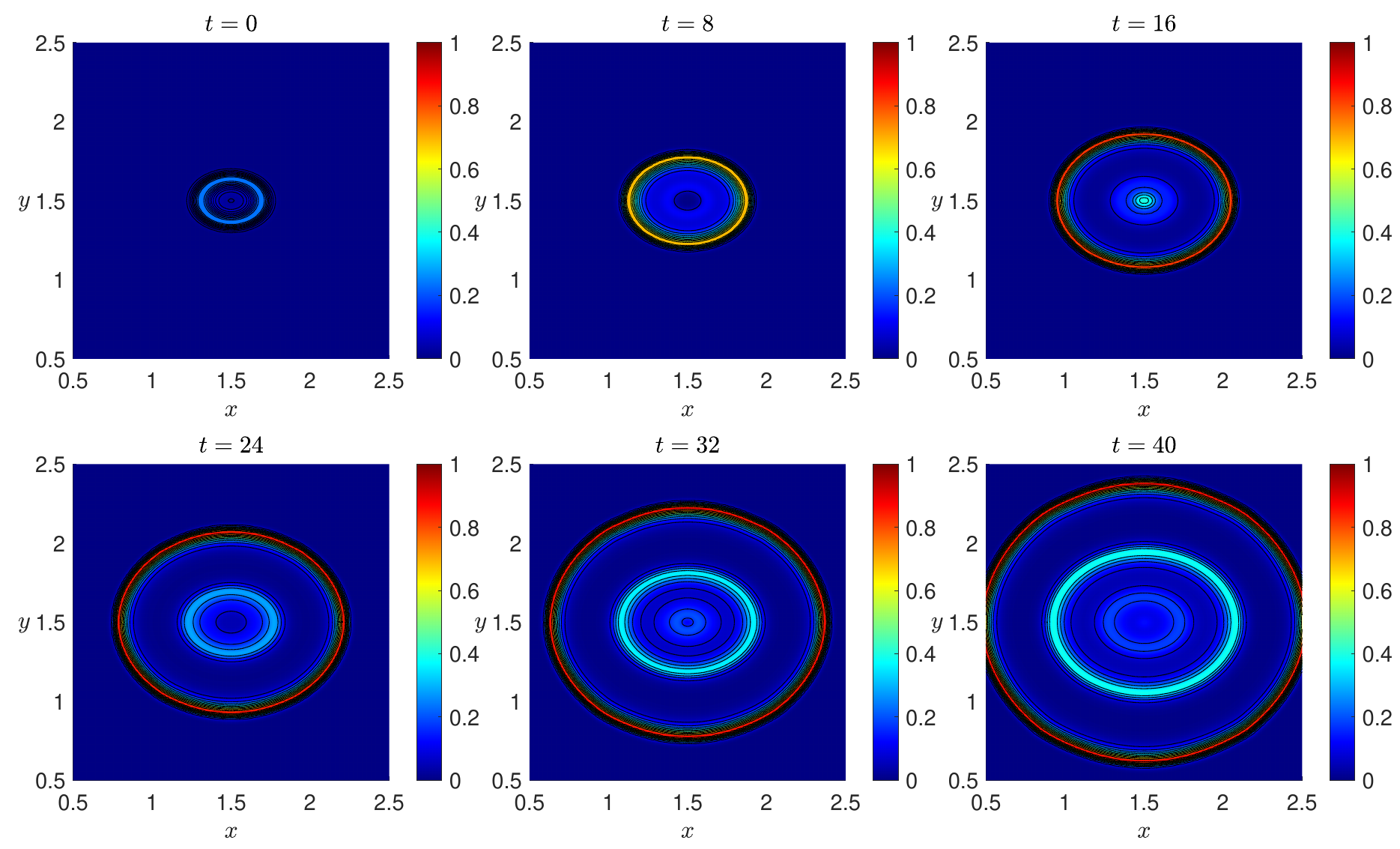}
\end{center}
\caption{Generation of quasi-periodic waves in the two-dimensional SIR model \eqref{eq:main:model:2D:vital} with cross-diffusion and vital dynamics.}
\label{fig:2D:wavegeneration}
\end{figure}

For the two-dimensional model \eqref{eq:main:model:2D:vital} with vital dynamics, similarly to the 1D case (Section \ref{sec3_1}), quasi-periodic infection waves arise. For an axially symmetric case, due to the autonomous nature of the PDEs \eqref{eq:main:model:2D:vital}, independence of the polar angle can be imposed, and the system can be reduced to a 1+1-dimensional model where $S$, $I$, and $R$ are functions of $(r,t)$, and $r$ is the cylindrical radius. In this case, the Laplacian in \eqref{eq:main:model:2D:vital} is given by $\Delta = (1/r)\partial/\partial r( r \partial/\partial r (\cdot))$. The same phenomenon of quasi-periodic wave production, however, also arises in the purely two-dimensional system. As an illustration, consider a spatial domain $V=[0,L]\times[0,L]$, parameters $L=3$, $D=0.10^{-5}$, $\gamma=0.4$, $\beta=1$, $\mu=0.1$, $m=0.9$, $K=4$, and $r=0.3$. For the initial condition $S_0=1$, $I_0 = \exp(-100(x-L/2)^2-200(y-L/2)^2)$, and $R_0=0$. The system dynamics is shown in Figure \ref{fig:2D:wavegeneration} where the formation of elliptic rings of decreasing amplitude originating from the initial infected population spike at the center of the domain is shown. The behaviour is parallel to that shown in Figure \ref{fig:3:1}.

%============================================================

\section{Discussion} \label{sec:disc}

In this work, our primary focus was the study of the spatiotemporal SIR model \eqref{m7} with spatial cross-diffusion terms, derived in one dimension in Ref.~\cite{bib2}. The nonstandard diffusion term is similar to that of the Keller-Siegel chemotaxis model \eqref{eq:KSiegel}; in the context of infectious disease modeling, it corresponds to infection propagation in the case of orderly commute of individuals moving to specified locations (``work") and returning to the original location (``home"). Extensions of this model proposed in this work include the derivation of the PDE system in two spatial dimensions and the study of effects of vital dynamics involving nonzero removal rates and logistic growth of the susceptible population (Section \ref{sec:2Dmodel}). In particular, the two-dimensional extension offers a natural setup to model a realistic population, whereas the logistic term is common in biological modeling where it is used, for example, to mimic internal competition for finite resources.

In Section \ref{sec2_1}, it is demonstrated that the PDE model \eqref{m7}, while being essentially non-hyperbolic, supports traveling waves of variable shape. Depending on system parameters, in particular, the diffusion coefficient $D$, these waves can accelerate or decelerate (Section \ref{sec2_1}).

In Section \eqref{sec4}, Lie point symmetry analysis of the PDE system \eqref{m7} is performed; it reveals unusual symmetries generated by $X_3$ and $X_4$ \eqref{eq:main:symm} that allow to add, to any solution of \eqref{m7}, a special Fourier mode-type time-decaying term
$\sin \left({\sqrt{\beta}x}/{\sqrt{a}}  \right)e^{-\gamma t}$ or $\cos \left({\sqrt{\beta}x}/{\sqrt{a}}  \right)e^{-\gamma t}$. This mode does not correspond to separation of variables (the PDEs \eqref{m7} are not separable) but is the only term of this kind, depending on the system parameters. It allows to generate new solutions of the PDE system \eqref{m7} from known ones according to the transformation \eqref{eq:symm:transfX3}. Two additional symmetries $X_5$ and $X_6$ are shown to arise for the special case when the model has a reduced form \eqref{m17}. In that situation, the use of these scaling symmetries allows to construct self-similar solutions arising from a single ODE \eqref{m21} and satisfying physical initial and boundary conditions on the half-line $x\geq 0$. The resulting solutions correspond to infection waves propagating in the domain $x>0$ in the presence of an infection source located at the origin.

Section \ref{sec:buffer} illustrates, for the original one-dimensional PDE model \eqref{m7} without vital dynamics, the existence of ``buffer zones" protected from the infection, by low population density. The existence of the buffer zone postpones the infection progression in the domain behind it, effectively working as firebreaks that are used to protect forests from fire propagation.

Special properties of the model that, in addition to cross-diffusion, involves vital dynamics (removal and logistic growth) terms, given by \eqref{eq:1D:with:vital},  are considered in Section \ref{sec:VDyn}. It is demonstrated that a single initial infection spike in an otherwise homogeneous population can cause the appearance of quasi-periodic infection waves that originate from the same location. The initial size of the susceptible compartment controls the time of the emergence of the waves. Buffer zones can also be present, resulting in a substantial delay in the propagation of infection waves beyond the buffer zone. Further, in case of homogeneous initial conditions, it is observed that the model supports spiral sink equilibria dependent on logistic term parameters. In the case of large carrying capacity, the equilibrium trajectory naturally approaches the corresponding level curve of the Lotka-Volterra-type Hamiltonian conserved integral.

Section \eqref{sec:equil:tr} is devoted to an unusual feature of the cross-diffusion model with vital dynamics, namely, local instability of perturbations of a homogeneous equilibrium state and ``dark spike" formation in the susceptible compartment (Figure \ref{pic15}). In particular, dark spikes form in the zones where the susceptible density is set to be initially lower; those areas quickly become narrower as time goes on. There are similar but wider and more transient gaps in recovered and infected populations in the same locations; this corresponds to overall low population density there. Eventually, as the births in the $S$-compartment build up the susceptible population, the system tends to the infection-free equilibrium. The dark spikes are essentially self-formed buffer zones. Mathematically, their emergence is related to linear instability of high-wavelength perturbations of the nontrivial steady state \eqref{eq:ss:star} of the PDE system  \eqref{eq:1D:with:vital}. We also note that for the ODE \eqref{eq:vit:ODE:S} describing time-independent states \eqref{eq:1D:with:vital:ODEs} of the system \eqref{eq:1D:with:vital}, numerical calculations indicate the existence of physically relevant, positive definite non-harmonic periodic solutions.

In Section \ref{sec:2D:study} it is demonstrated that, in a manner parallel to the one-dimensional case, the two-dimensional PDE model \eqref{m8} with cross-diffusion also features buffer zones protected from infection, and that the full system \eqref{eq:main:model:2D:vital} with vital dynamics features the generation of self-induced quasiperiodic infection waves propagating across the population (Figure \ref{fig:2D:wavegeneration}; cf.~Figure \ref{fig:3:1}).

In future work dedicated to cross-diffusion models considered in this work, it is essential to further explore essential features of the models, in particular, wave-type phenomena that arise for this set of non-hyperbolic diffusion-type nonlinear autonomous PDEs. It is of interest to analyze the wave propagation speeds and wave shapes in one-dimensional dynamics; derive, perhaps in a certain asymptotic limit, the time period of wave generation in models with vital dynamics; analyze types of solutions admitted by the time-independent nonlinear ordinary differential equations \eqref{eq:1D:with:vital:ODEs}, \eqref{eq:vit:ODE:S}; provide a better biological interpretation of the formation of transient dark spikes; further improve the PDE system from the point of view of biological modeling.

\section*{Acknowledgments}
A. C. thanks NSERC of Canada for research support through a Discovery grant RGPIN-2024-04308.

\section*{Competing interests declaration}
 No competing interests are involved in any form.

\bibliographystyle{ieeetr}
{\small
\bibliography{CA}
}

\end{document}